\documentclass[conference]{IEEEtran}

\usepackage{graphicx}
\usepackage{dcolumn}
\usepackage{bm}
\usepackage{amsmath,amsfonts,amscd,revsymb,latexsym, enumerate,multirow}
\usepackage[numbers,sort&compress]{natbib}

\def\bu{{\bf u}}

\def\bx{{\bf x}}

\begin{document}
%
\title{Performance of Polar Codes on wireless communications Channel}
\author{\IEEEauthorblockN{Peng Shi}
\IEEEauthorblockA{Institute of Signal \\
Processing \&\\Transmission\\
Nanjing University of Posts\\
\& Telecommunications\\
Nanjing, China}
\and
\IEEEauthorblockN{Shengmei Zhao}
\IEEEauthorblockA{Institute of Signal \\
Processing \&\\Transmission\\
Nanjing University of Posts\\
\& Telecommunications\\
Nanjing, China}
\and
\IEEEauthorblockN{Bei Wang}
\IEEEauthorblockA{Institute of Signal \\
Processing \&\\Transmission\\
Nanjing University of Posts\\
\& Telecommunications\\
Nanjing, China}
\and
\IEEEauthorblockN{Liang Zhou}
\IEEEauthorblockA{Institute of Signal \\
Processing \&\\Transmission\\
Nanjing University of Posts\\
\& Telecommunications\\
Nanjing, China}}
\maketitle
\begin{abstract}
\boldmath
We discuss the performance of polar codes, the capacity-achieving channel codes, on wireless communication channel in this paper. By generalizing the definition of Bhattacharyya Parameter in discrete memoryless channel, we present the special expression of the parameter for Gaussian and Rayleigh fading the two continuous channels, including the recursive formulas and the initial values. We analyze the applications of polar codes with the defined parameter over Rayleigh fading channel by transmitting image and speech. By comparing with low density parity-check codes(LDPC) at the same cases, our simulation results show that polar codes have better performance than that of LDPC codes. Polar codes will be good candidate for wireless communication channel.
\end{abstract}
\IEEEpeerreviewmaketitle
\section{Introduction}
\boldmath
Polar codes, proposed by Arikan, are the first provably capacity-achieving family of codes with low encoding and decoding complexity \cite{Arikan2007channel}.  They have attracted much attention recently \cite{Arikan2007channel,sasoglu2009polarization,korada2010polarsource,Karzand2010Polar, Korada2010Polar,presman2011polar, Abbe2010MAC, Mori2010Non,
Blasco2010Polar,Maha2010achieving,Mark2011Cla,Mark2011deg,Arikan2008A}.
After the channel polarization which is the key technique for polar codes has been proposed, polar codes have been generalized both in definition and application.
In \cite{sasoglu2009polarization}, polar codes are generalized to the case
of memoryless output-symmetric channels with non-binary input alphabets. In \cite{korada2010polarsource,Karzand2010Polar}, polar codes are extended to source coding for achieving the rate-distortion bound in lossy source compression.  In
\cite{Korada2010Polar,presman2011polar}, the polarization phenomenon is studied for a broad family of kernel matrices and error exponents are derived for each such kernel matrix.
In terms of applications, polar codes
have been used in the context of multiple access channels \cite{Abbe2010MAC}, relay channels \cite{Blasco2010Polar}, wiretap channels \cite{Maha2010achieving}, and quantum channels \cite{Mark2011Cla,Mark2011deg}.
\par
Comparing with the theoretic proofs on polar codes, the performance of polar codes are relatively less analyzed, and they are mainly discussed on discrete channels, such as, binary memoryless channels(BMC)\cite{Arikan2008A,Mori2010Non}, binary erasure channel(BEC)\cite{Arikan2008A}. A more practical communication channel, the wireless communications channel, is so far less concerned. 
\par
The typical wireless communications channels are Gaussian channel (AGNC) and Rayleight channel. They are continuous channels. In this paper, we focus on the performance of polar codes on  the continuous communication channels. First, we present the formalism of Bhattacharyya parameter for Gaussian and Rayleigh channel. 
Then, we construct polar codes on these two kinds of channels. Finally, we present some numerical experiments of polar codes with image and speech transmission over Rayleigh channel, and compare the results with those of low density parity check codes (LDPC), the good candidate codes on currently wireless communications channel.
\section{Background }
\boldmath
 It is the unique construction method, named channel polarization, makes polar codes having the capacity achieving property. By channel polarization, the bit-channels will divide into noiseless bit-channels or pure-noise bit-channels.   In polar codes design, only those noiseless bit-channels are selected to transmit information. 
As a family of block codes, polar codes can be described as
\begin{equation}
\label{eqn_1}
x_1^N  = u_1^N G_N,
\end{equation}
where $N$ is the block length and is always set to $2^n$, $n \ge 0$. $G_N$ is the generator matrix of polar codes,  which comes from the $n$-th Kronecker power of $G_2$. 
The standard matrix of $G_2$ is
\begin{equation}
G_2  = \left[ {\begin{array}{*{20}c}
   1 & 0  \\
   1 & 1  \\
\end{array}} \right].
\end{equation}
  Polar codes can be defined by three parameters: block-length $N$, rate $R=K/N$, and an information set $A$, denoted as ($ N $,$ K $,$ A $,$\bu_{A^C}$). Here, $\bu_{A^C}$ is `frozen' bits. They are kept constant and the decoder can avoid errors in the frozen part.  A polar codeword $\bx$ may be obtained by mapping a binary $N$-tuple $\bu$ whose $i$th component is set (`frozen') to zero for all $i \in \left[N\right] /A$. Hence, $\left(\ref{eqn_1}\right)$ can also be written as
\begin{equation}
x_1^N  = u_A G_N (A) \oplus u_{A^C } G_N (A^C ),
\end{equation}
where $G_N$($A$) is a submatrix of $G_N$ formed by the rows with indices of $A$. Undoubtedly,  polar codes are channel-specific designs.  A construction method of polar codes for one channel may be not suitable for another. 
\par
The noiseless or pure-noise of bit-channels is determined by the Bhattacharyya parameter $Z(W)$ of actual channel $W$. For any discrete memoryless channel (DMC) channel $W$, $Z(W)$ is defined as
\begin{equation}
\label{eqn_2}
Z(W) = \sum\limits_y {\sqrt {W(y|0)W(y|1)} },
\end{equation}
and the block error probability of polar codes is proved to
\begin{equation}
P_e (N,K,A,u_{A^c } ) \le \sum\limits_{i \in A} {Z(W_N^{(i)} )}.
\end{equation}
\section{designs of Bhattacharyya parameter for continuous channel}
The Bhattacharyya parameter is defined as the parameter to measure the reliability of channel. It is obvious that the definition of $Z(W)$ in (\ref{eqn_2}) is not suitable for the continuous channel. However, it is reasonable to expand this definition from discrete channel to continuous channel because of the similarity of the two kinds channels. As the extension from discrete channel, we use the density of probability distribution instead of probability distribution, and the integrating instead of summing for the Bhattacharyya parameter in continuous channel. Hence, the formulation can be expressed as
\begin{equation}
Z(W) = \int {\sqrt {W(y|0)W(y|1)} } dy.
\end{equation}
\boldmath
Then, the recursive properties of channel polarization can be calculated by the iteration of the Bhattacharyya parameter with
\begin{equation}
\begin{split}
 Z(W_N^{(i)} ) & \le Z(W_{2N}^{(2i - 1)} )  \\
& \le 2Z(W_N^{(i)} ) - Z(W_N^{(i)} )^2, \\
 Z(W_{2N}^{(2i)} ) & = Z(W_N^{(i)} )^2.
\end{split}
\end{equation}
If the channel $W$ is a BEC, the first formula can be written as\cite{Arikan2007channel}
\begin{equation}
Z(W_{2N}^{(2i - 1)} ) = 2Z(W_N^{(i)} ) - Z(W_N^{(i)} )^2.
\end{equation}
However, it is more complicated for AGNC and Rayleigh channel. 
 In order to obtain the suitable recursive relation of Bhattacharyya parameters for continuous channel, we present the following three types iterations according to the definition, and choose a better one for the next section numerical experiments. The three types recursive relations of Bhattacharyya parameters are as follows
\begin{equation}
\begin{split}
 Type 1: & Z(W_{2N}^{(2i - 1)} ) = 2Z(W_N^{(i)} ) - Z(W_N^{(i)} )^2, \\
 Type 2: &  Z(W_{2N}^{(2i - 1)} )=  Z(W_N^{(i)} ), \\
 Type 3:  & Z(W_{2N}^{(2i - 1)} ) = 0.5 \times (2Z(W_N^{(i)} ) \\
            &   -Z(W_N^{(i)} )^2  + Z(W_N^{(i)} )). \\
\end{split}
\end{equation}


\begin{figure}[hb]
\centering
\includegraphics{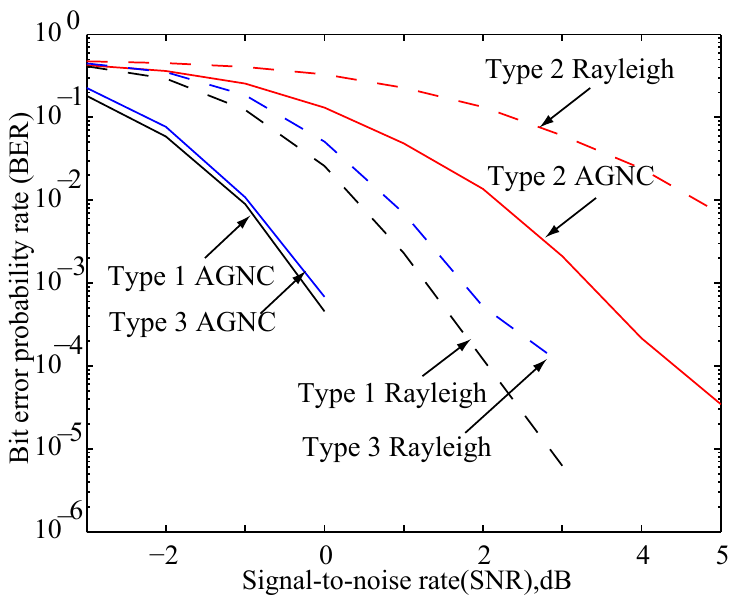}
\caption{The comparison of three types recursive relations of Bhattacharyya parameters on AGNC and Rayleigh channel with $N$=256 and $R$=0.25. The bold lines represent the results for AGNC channel  and the dot lines represent the results for Rayleigh channel.}
\label{fig1}
\end{figure}
\par
  Fig.~\ref{fig1} shows the comparison of the three types recursive relations of the Bhattacharyya parameter on AGNC and Reyleigh channel, the abscissa shows the signal-to-noise rate, the vertical axis shows the bit error rate probability. The bold lines  and the dot lines represent the results for AGNC and Rayleigh channel with $R$=0.25 respectively. We use 10000 frames for each simulation, and the code block length is 256. The results show that
 'Type1' has a better performance among these three kinds recursive relations.
\par
To calculate the Bhattacharyya parameter, the next key element is the initial value. In \cite{Arikan2008A}, the initial value of the Bhattacharyya parameter $Z_0$ is set to 0.5 for the channels. Absolutely, it is not a good one for the continuous communication channels. We now discuss the best initial value $Z(W)$ for AGNC and Rayleigh channel.
\par
Let's consider AGNC channel first. Suppose there is a communication link with Gaussian noise with expectation $0$ and variance  $\sigma ^2$. Meanwhile, phase-shift key (PSK) modulation is adopted as modulation. Then, the output of AGNC is a signal with expectation $1$ or $-1$ and variance $\sigma ^2$. Therefore, the density probability for sending $0$ ( or $1$) symbol and getting the output $y$ is  $W(y|0)$ ( or $W(y|1)$). They are
\begin{equation}
\begin{split}
W(y|0) & = \frac{1}{{\sqrt {2\pi } \sigma }}e^{ - \frac{{(y + 1)^2 }}{{2\sigma ^2 }}}, \\
W(y|1) & = \frac{1}{{\sqrt {2\pi } \sigma }}e^{ - \frac{{(y - 1)^2 }}{{2\sigma ^2 }}}. \\
\end{split}
\end{equation}
  We can use them as the initial value for AGNC channel.
\par
In comparison with AGNC channel, Rayleigh channel has  more complicated initial value of the Bhattacharyya parameter.
As we know, Rayleigh fading  is normally described as
\begin{equation}
a = K\sqrt {x^2  + y^2 },
\end{equation}
where $x$, $y$ are both the independent Gaussian random variables with expectation $0$ and variance $1$, and $K$ is often called scaling factor. 
For simplicity, we assume that the expectation $E$ is the peak value of Rayleigh distribution, and the variance is connected with signal-to-noise ratio(SNR) $S$. Based on Rayleigh distribution's properties, we can calculate that the expectation is $\pm \sqrt {2\ln 4} K$ and variance $\sigma ^2$ is  $\left(10^{\frac{S}{10}}\right)^{-1}$
respectively. Therefore, $W(y|0)$ and $W(y|1)$ can be written as,
\begin{equation}
W(y|0) = \frac{1}{{\sqrt {2\pi } \sigma }}e^{ - \frac{{(y + \sqrt {2\ln 4} K)^2 }}{{2\sigma ^2 }}},
\end{equation}
and
\begin{equation}
W(y|1) = \frac{1}{{\sqrt {2\pi } \sigma }}e^{ - \frac{{(y - \sqrt {2\ln 4} K)^2 }}{{2\sigma ^2 }}}.
\end{equation}
In order to guarantee the average power of Rayleigh fading to be 1, we set $K$ to 0.7 in the later simulations.
\begin{figure}[hb]
\centering
\includegraphics[width=3.5in]{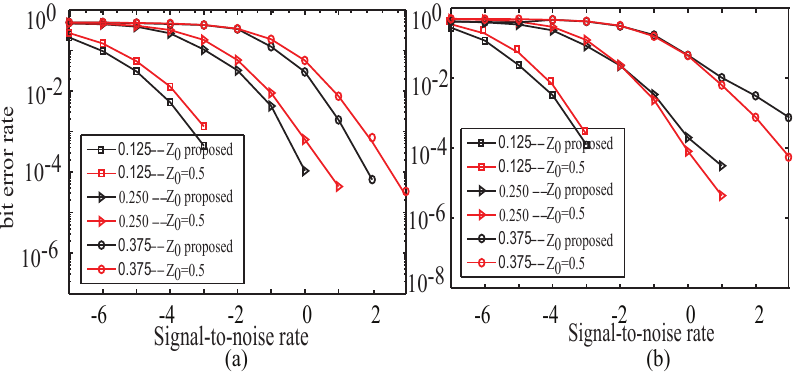}
\caption{The bit error probability rate (BER) of polar codes for wireless communications channel with proposed and constant Bhattacharyya parameter initial value, where (a) is on AGNC channel and (b) is on Rayleigh fading channel.}
\label{fig2}
\end{figure}
\par
Fig.~\ref{fig2} shows the comparison results with proposed and constant Bhattacharyya parameter initial value on AGNC and Rayleigh channel. Fig.~\ref{fig2}(a) presents the bit error rate (BER) performance on AGNC channel, and Fig.~\ref{fig2}(b) presents the results on Rayleigh channel. The code rates $R$ are 0.125, 0.25 and 0.375 respectively. It's obvious that the BER performance of polar codes using proposed $Z_0$ is lower than that with constant initial value $Z_0$=0.5 for AGNC. However, the results are some different for Rayleigh channel. The BER performance using proposed $Z_0$ is better than that with $Z_0$=0.5 when signal-to-noise (SNR) is lower than 1dB, while $Z_0$=0.5 has a better performance when SNR is greater than 1dB. Therefore, we could use the proposed $Z_0$ for AGNC channel. For Rayleigh channel, we could set initial value $Z_0$ to be 0.5 when SNR is greater than 1dB, and use proposed $Z_0$ when  SNR is lower than 1dB.
\section{numerical simulations on Rayleigh channel}
\begin{figure}[htb]
\centering
\includegraphics{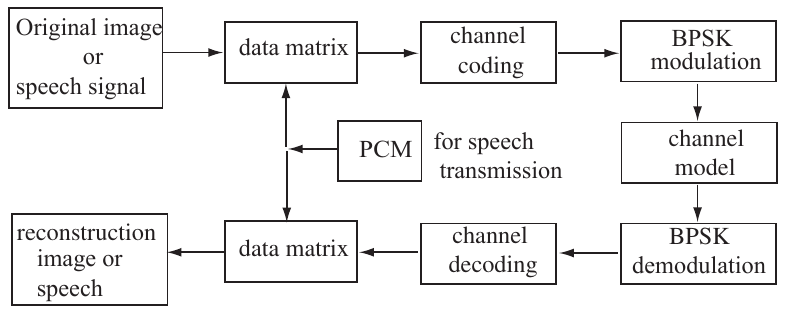}
\caption{The transmission model of image and speech over wireless communications channel. The data matrix module is replace by PCM module in speech transmission. }
\label{fig3}
\end{figure}

In this section, we will apply polar codes to wireless communications channel with image and speech transmission and compare the results with that with Low density parity check (LDPC) codes. For simplicity, we only discuss the transmissions on Rayleigh channel. 
 LDPC codes which were introduced by Gallager\cite{Gallager1962Low,Yuling2004the,Wei2009Per} are nowadays used capacity-achieveing channel codes. They are found widely applications in wireless communications channel. We compare the performance of image and speech transmission system using polar codes with using LDPC codes. Both comparisons are based on the transmission model shown in Fig.~\ref{fig3}.

\boldmath
\par
\subsection{Image transmission experiments}
\boldmath
First, we testify the image transmission with polar and LDPC coding. The size of original image is 256$\times$256 pixels with 256 grayscales. The channel noise is Rayleigh fading. Additionally, binary phase-shift keying modulation (BPSK) is selected as modulation. Normally, peak signal to noise (PSNR) is used as an objective evaluation for the reconstruction. We use the definition of PSNR as follows
\begin{equation}
\begin{split}
MSE(a,b) & = \frac{1}{{N \times M}} \\
 & \sum\limits_{x = 0}^{N - 1} {\sum\limits_{y = 0}^{M - 1} {\left[ {a(x,y) - b(x,y)} \right]} ^2 }, \\
PSNR(a,b) & = 10\lg \left[ {\frac{{255^2 }}{{MSE(a,b)}}} \right]. \\
\end{split}
\end{equation}
where $a(x,y)$ is the grayscale of original image at $(x,y)$,  $b(x,y)$ is the reconstruction grayscale, $M$ and $N$ are two dimensional sizes of the image. It demonstrates that the greater PSNR is, the better of the reconstruction image is.
\begin{figure}[htb]
\centering
\includegraphics{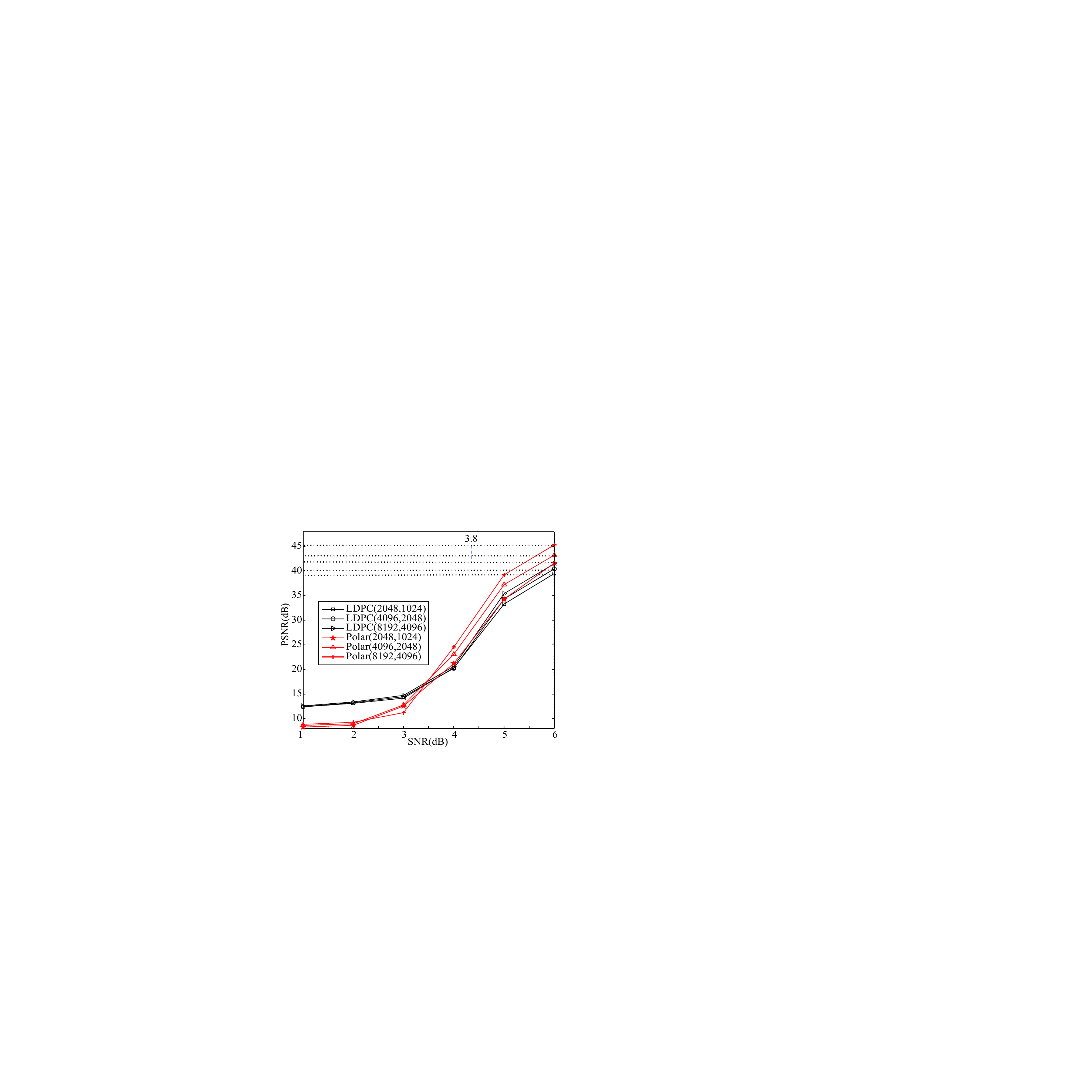}
\caption{ The PSNR of image transmission with polar codes and LDPC codes over Rayleigh channel.
}
\label{fig4}
\end{figure}
\par
\par
Fig.~\ref{fig4} shows the PSNR of reconstruction image with polar codes and LDPC codes over Rayleigh channel. Each value at the figure is the average over 1000 times. The code lengths of polar codes and LDPC codes are 2048, 4096 and 8192 respectively and the code rate is set to $0.5$. When the SNR of Rayleigh channel is 6dB, the PSNR of polar codes is greater than that of LDPC by 1dB with code length being 2048. The PSNR difference for the two codes increases to 3.8dB when the code length is 8192. Furthermore, the performance of polar codes improves more greatly than that of LDPC codes with the increase of code length.
\begin{figure}[htb]
\centering
\includegraphics{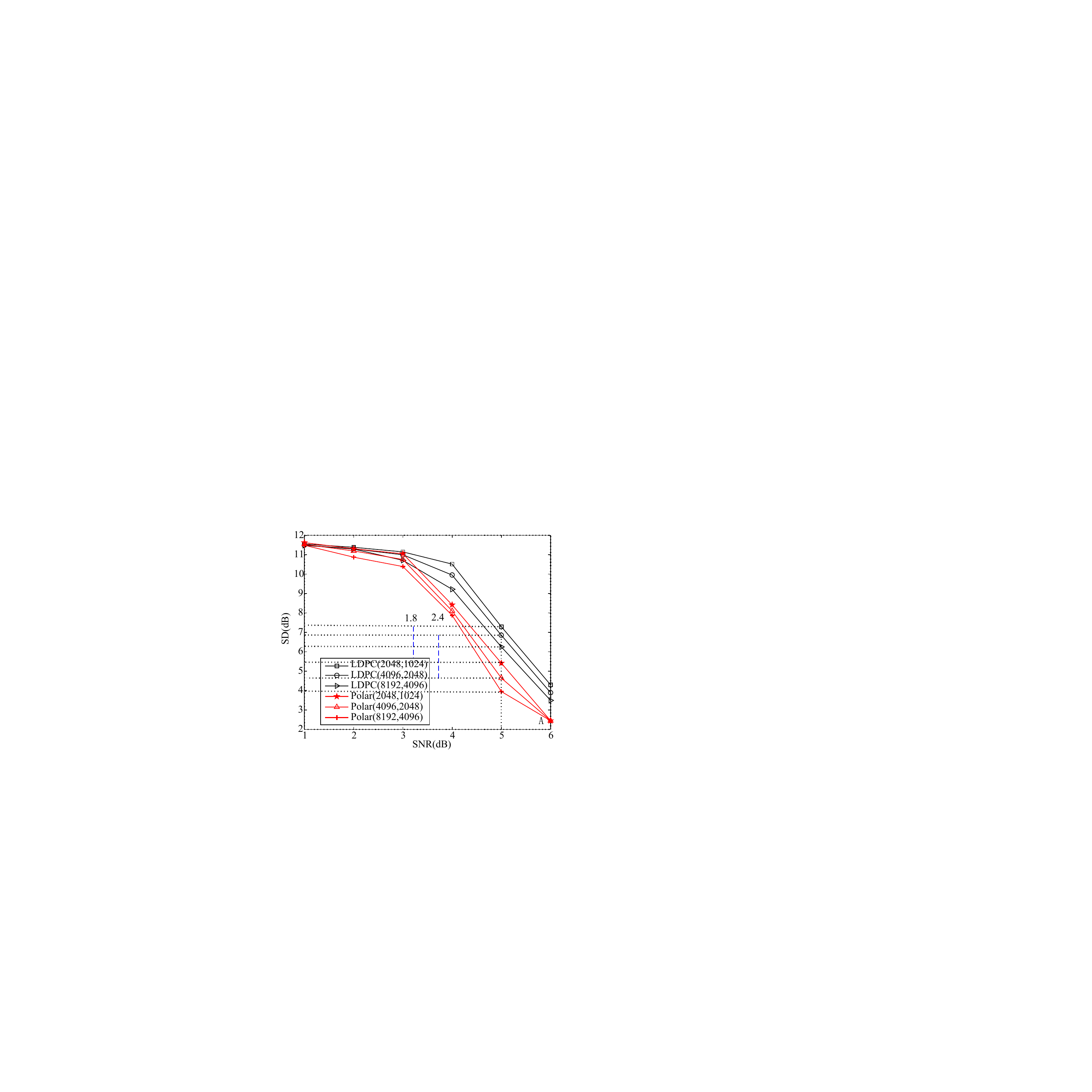}
\caption{The SD of different channel codes for Rayleigh channel on the speech communications system. }
\label{fig5}
\end{figure}
\subsection{Speech transmission experiments}
\boldmath
Then, we discuss the performance of polar codes and LDPC codes on the speech communications system. It is known that channel coding can be also adopted to improve the speech communications quality by information redundancy. At this time, pulse-code modulation (PCM) will be used as sampling method to collect the data matrix from continuous signal to discrete ones in Fig.~\ref{fig3}. In the experiments, the source speech has 5320 frames. The average spectral distortion  ($SD$) is adopted as the objective evaluation parameter for speech waveform coding, which is defined as
\begin{equation}
\begin{split}
SD & = \frac{1}{{N}}\sum\limits_{j = 1}^N[\int\limits_{ - \pi }^\pi  {(10\log _{10} S_j (\omega )}\\
 &{- 10\log _{10} S'_j (\omega ))^2 \frac{{d\omega }}{{2\pi }}]^{1/2} },
\end{split}
\end{equation}
where $S_j(\omega)$ denotes the original speech spectral associated with the frame $j$, while $S'_j (\omega )$ denotes the reconstruction speech spectral. And the lower $SD$ is corresponding to the better reconstruction.
\par
Fig.~\ref{fig5} is the experimental result of reconstructed $SD$ coded by polar codes and LDPC codes on Rayleigh channel. The codes lengths are 2048, 4096 and 8192 respectively and the code rate is 0.5. In the figure, $SD$=2.45dB (point A) is the situation without any transmission error by PCM modulation. The numerical results show that polar codes have better performance than LDPC codes for the speech communications system over Rayleigh channel. When the SNR is 5dB and the code length is 2048, the $SD$ of polar codes is 5.4dB, while that of LDPC is 7.2dB. The performance of polar codes improves more greatly than that of LDPC with the increase of code length. For example, when the code length is twice (4096), the $SD$ difference for the two codes increases from 1.8dB to 2.4dB with SNR being 5dB.
\section{Conclusion}
 In this paper, the performance of polar codes on AGNC and Rayleigh channel, the two normally wireless communications channel, has been discussed. At first, the special Bhattacharyya Parameter expression for the two continuous channels have been defined, including the recursive formulas and the initial values. With these definitions, the applications of polar codes over Rayleigh fading channel have been discussed by transmitting an image and a speech sample. The results have been compared with those from the system with LDPC codes at the same conditions. For the image transmission, polar codes have a better performance than LDPC codes when PSNR is greater than 3.5dB. For the speech transmission, polar codes can have a better reconstruction speech. In addition, with the increase of code length, the performance of polar codes in image and speech transmission improves more quickly than that of LDPC codes. All the results show polar codes have a better performance than that of LDPC codes over Rayleigh channel, and they will be a good choice for wireless communications channel.

\section*{Acknowledgment}
The work of this paper was supported in part by University Natural Science Research Foundation of JiangSu Province (11KJA510002),  the Foundation (No.NJ210002),  the open research fund of Key Lab of Broadband Wireless Communication and Sensor Network Technology, Ministry of Education, and the project funded by the Priority Academic Program Development of Jiangsu Higher Education Institutions.


\begin{thebibliography}{99}
\bibitem{Arikan2007channel}Arikan, E., 'Channel Polarization: A method for constructing capacity-achieving codes for symmetric bianry-input memoryless channels', IEEE Trans. Inf. Theory \textbf{55}, 3051-3073, 2007

\bibitem{sasoglu2009polarization}Sasoglu, E., Telatar, E. and Arikan, E., Polarization for arbitrary discrete memoryless channels, in 'IEEE Info. Theory Workshop(ITW)', pp. 144-148, 2009

\bibitem{korada2010polarsource}Korada, S. and Urbanke, R., 'Polar codes are optimal for lossy source coding', IEEE Trans. Inf. Theory 56(4), 1751-1768, 2010

\bibitem{Karzand2010Polar}Karzand, M. and Telatar, E., Polar codes for Q-ary source coding, in 'In Proc. IEEE Int. Symp. Information Theory (ISIT)', pp. 909-912, 2009

\bibitem{Korada2010Polar}Korada, S.; Sasoglu, E. and Urbanke, R., 'Polar codes: Characterization of exponent, bounds, and constructions', IEEE Trans. Inf. Theory \textbf{56}, 6253-6264, 2010

\bibitem{presman2011polar}Presman, N.; Shapira, O. and Litsyn, S., Polar codes with mixed kernels, in 'In Proc. IEEE Int. Symp. Information Theory (ISIT)', pp. 6-10, 2011

\bibitem{Abbe2010MAC}Abbe, E. and Telatar, E., MAC Polar codes and matroids, in 'In Proc. Workshop on Information Theory and Applications (ITA)', pp. 1-8, 2010

\bibitem{Mori2010Non}Mori, R. and Tanaka, T., Non-binary Polar codes using Reed-Solomon codes and algebraic geometry codes, in 'IEEE Trans. Info. Theory Workshop(ITW)', pp. 1-5, 2010

\bibitem{Blasco2010Polar}Blasco-Serrano, R.; Thobaben, R.; Rathi, V. and Skoglund, M., Polar codes for compress-and-forward in binary relay channel, in 'ASILOMAR 2010', pp. 1743-1747, 2010

\bibitem{Maha2010achieving}Mahdavifar, H. and Vardy, A., Achieving the secrecy capapcity of Wiretap channels using Polar codes, in 'In Proc. IEEE Int. Symp. Information Theory (ISIT)', pp. 913-917, 2010

\bibitem{Mark2011Cla}Wilde, M. M. and Guha, S., 'Polar codes for classical-quantum channels', arXiv:1109.2591v2 [quant-ph], 2011

\bibitem{Mark2011deg}Wilde, M. M. and Guha, S., 'Polar codes for degradable quantum channels', arXiv:1109.5346v2 [quant-ph], 2011

\bibitem{Arikan2008A}Arikan, E., 'A performance Comparison of Polar codes and Reed-Muller Codes', IEEE Comm. Letters 12, 3447-449, 2008

\bibitem{Gallager1962Low}Gallager, R. G., 'Low density parity check codes', IRE, Trans. Info. Theory 8, 21-28, 1962

\bibitem{Yuling2004the}Zhang, Y.; Yuan, D.; Zhang, H. and Zhang, H., The Application of LDPC Codes In Image Transmission, in 'Communications, Circuits and Systems(ICCA)', pp. 29-33, 2004

\bibitem{Wei2009Per}Han, W.; Huang, J. and Jiang, M., Performance Analysis of Underwater Digital Speech Communication System Based on LDPC codes, in 'Industrial Electronics and Applications(ICIEA)', pp. 567-570, 2009

\end{thebibliography}
\end{document}